# Hole content dependent fluctuation diamagnetism in $YBa_2Cu_3O_{7-\delta}$: possible role of the pseudogap

*Ayesha Siddika Borna, R. S. Islam, and S. H. Naqib\**

*Department of Physics, University of Rajshahi, Rajshahi 6205, Bangladesh*

*\*Corresponding author: salehnaqib@yahoo.com*

**Abstract**

This study focuses on the temperature and hole content dependent fluctuation diamagnetism of hole doped $YBa_2Cu_3O_{7-\delta}$ (Y123) high-$T_c$ superconductors. Two different compositions of Y123 have been considered with in-plane hole content (p): 0.161 (optimally doped) and 0.143 (underdoped). The fluctuation induced excess diamagnetic susceptibility, $\Delta\chi(T)$, has been investigated via the mean-field Gaussian-Ginzburg-Landau (MFGGL) formalism with and without a total energy cut-off in the fluctuating modes. It has been found that inclusion of total energy cut-off describes the $\Delta\chi(T)$ data significantly better. Furthermore, the pseudogap (PG) itself induces an anomalous decrease in the normal state magnetic susceptibility. By means of the analysis of $\Delta\chi(T)/T$ at different hole concentrations, we have explored the possible role of the PG on diamagnetic fluctuations. It has been found that MFGGL formalism is not able to reproduce the $\Delta\chi(T)/T$ features for the underdoped compound over a broad range of reduced temperature, $\varepsilon$ [= $\ln(T/T_c)$]. The discrepancy becomes prominent in the temperature range where PG dominates the normal state magnetic susceptibility data. The agreement between the theoretical prediction and experimental $\Delta\chi(T)$ is better for the optimally doped compound with p = 0.161, where the effect of the PG is small. This notable difference implies that PG induced reduction in the magnetic susceptibility is not related directly to the superconducting fluctuations which in turn indicate that electronic correlations giving rise to the PG and Cooper pairing are independent to each other.

## 1. Introduction

Although electron pairing above the superconducting (SC) transition temperature, $T_c$, is not favored because of the higher free energy, there are always short-lived pairing fluctuations. These fluctuating Cooper pairs, at temperatures above $T_c$, decrease the magnetic susceptibility in the normal state. This is known as the fluctuation diamagnetism.

Since the early days, the effects of SC fluctuations on electrical conductivity (paraconductivity) and magnetic susceptibility in the normal state have been studied extensively for variety of conventional BCS (Bardeen-Schrieffer-Cooper) superconductors [1 – 4]. The effect of Cooper pairing fluctuations in these low-$T_c$ systems can be understood within the time-dependent Ginzburg-Landau formalism [5, 6]. The situation for hole doped high-$T_c$ cuprates is different. Copper oxide superconductors exhibit much stronger thermodynamic fluctuations than the conventional low-$T_c$ ones because of their higher $T_c$, shorter SC coherence lengths and quasi-two-dimensional crystal structures [5, 7 – 9]. Furthermore, experimental observations of diamagnetic signals and substantial Nernst coefficients over a broad temperature range well above the SC critical temperatures in different families of hole doped cuprates [10, 11] are fascinating and are often put forward as evidences for preformed Cooper pairs without the long-range phase coherence which is needed for superconducting order. On the other hand, the presence of the PG in underdoped (UD) and optimally doped (OPD) copper oxide superconductors progressively reduces both resistivity and magnetic susceptibility in the normal state [12 - 15], the decrement of normal state resistivity and magnetic susceptibility with decreasing temperature due to SC fluctuations near $T_c$ is hard to distinguish from the decrement due to the presence of the PG. This creates a significant theoretical challenge in understanding the paraconductivity and fluctuation diamagnetism in high-$T_c$ cuprates. The problem can be envisaged from two different angles, both connected to the possible origin of the PG in the low energy quasiparticle spectral density [16 – 18] in hole doped cuprates.

The origin and evolution of the PG in high-$T_c$ cuprates with temperature and hole content are hotly debated issues. Broadly speaking, two different schools of thought exist [16 – 18]. The first one is based on the precursor Cooper pairing scenario. In this scheme, electrons form phase incoherent pairs at a temperature T* which can be much larger than the critical temperature $T_c$ where phase coherent superconductivity appears. In this picture, superconductivity does not appear at T* because large phase fluctuation of the pairing field cannot order at such high temperature [16 – 19]. The pseudogap is then produced by incoherent fluctuations of the pairing field [19]. As hole content (p) increases, T* falls and eventually merges with $T_c$ in the overdoped side of the electronic phase diagram. In the second scenario, the PG is attributed to some correlations of non-SC origin and PG coexists and in fact, competes with superconductivity [12, 13, 16, 20]. In this scenario T*(p) also decreases with increasing hole content but never merges with $T_c$(p), instead crosses the parabolic $T_c$(p) line at around the optimum doping (p = 0.16) and goes below the $T_c$(p) line and vanishes at p ~ 0.19 [12 – 15, 20, 21]. The schematic behaviors of these two scenarios are shown in Fig. 1 [16 – 18]. Within the first scenario, the PG induced downturns in the resistivity and magnetic susceptibility have their origin in the phase incoherent pairing fluctuations and, hence, should be accountable to the Gaussian-Ginzburg-Landau (GGL) type mean field fluctuation analysis above the SC transition temperature. In the second scenario the PG induced downturns in the normal state resistivity and magnetization are unrelated to the pairing correlations and, therefore, an GGL type pairing fluctuation analysis might be inadequate to account for the experimentally extracted paraconductive and fluctuation diamagnetic susceptibility contributions. In this study we wish to address this issue.

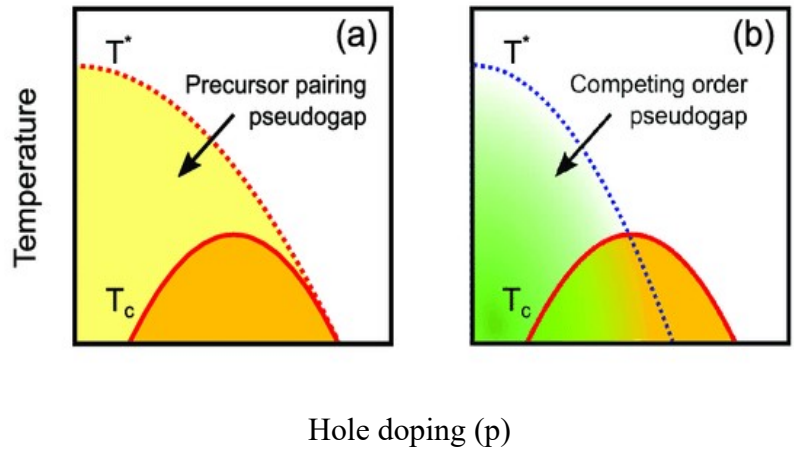

Figure 1: Schematic T-p phase diagram of hole doped high-$T_c$ cuprates in the (a) precursor pairing scheme of the pseudogap and (b) independent correlations picture.

The earliest experimental work on the fluctuation induced diamagnetism above $T_c$ in superconductor was reported by Gollub, Tinkham and co-workers, over half a century ago [2, 22, 23]. All these studies involve different bulk isotropic metallic superconductors with low $T_c$. In cuprates, fluctuation diamagnetism was studied first qualitatively by Freitas et al. [24] and quantitatively later by Lee et al. [25]. Since then a significant body of work has been accumulated [9, 26 – 29] which is devoted on the fluctuation diamagnetism of hole doped cuprates belonging to different families. Very recently, Rey et al. [30] have studied the paraconductivity and fluctuation diamagnetism of OPD $YBa_2Cu_3O_{7-\delta}$ single crystal keeping the possible role of the PG in mind. Rey et al. concluded that PG has no significant effect on the SC fluctuation of OPD $YBa_2Cu_3O_{7-\delta}$ and the observed paraconductivity and fluctuation diamagnetism both can be explained by GGL type mean field fluctuation analysis above the SC transition temperature [30]. This is not surprising, because the T*(p) of OPD $YBa_2Cu_3O_{7-\delta}$ is quite low – close to $T_c$ and the PG induced downturns in resistivity and magnetization are insignificant over a wide range of temperature above $T_c$. In other words, for the OPD compound the effect of the PG is minimal in the normal state to start with. To investigate and unentangle the possible role of PG and SC fluctuation with reliability, one needs to analyze the paraconductivity/fluctuation diamagnetism in cuprates where the PG is significant. A hole content dependent analysis is required so that the effect of the PG can be properly identified. Such studies are extremely scarce in the literature. In a previous investigation [31], we have analyzed the paraconductivity of high quality crystalline thin films of $YBa_2Cu_3O_{7-\delta}$ with different hole concentrations to address the role of the PG. Here we focus on the fluctuation diamagnetism of $YBa_2Cu_3O_{7-\delta}$ with different hole contents.

The remainder of the paper is structured as follows. Section 2 consists of experimental results and the analysis. In Section 3, we have discussed the implications of the analysis and drawn the main conclusions of this work.

## 2. Experimental results and analysis

### 2.1. Samples and characterizations

Polycrystalline single-phase sintered samples of $YBa_2Cu_3O_{7-\delta}$ were synthesized by standard solid state reaction method using high purity starting compounds. Samples were characterized by X-ray diffraction (XRD), low-field ac susceptibility (ACS), resistivity ($\rho(T)$), and room-temperature thermopower (S[290 K]) measurements. The hole content was varied by changing the oxygen deficiency ($\delta$) in the $CuO_{1-\delta}$ chains by quenching the samples into liquid nitrogen from different elevated temperatures and oxygen partial pressures. Further details regarding sample preparation and oxygen annealing can be found in Refs. [32, 33]. The magnetic susceptibility, $\chi(T)$, data for Y123 used in this study have been reported in previous studies [15, 33]. The room temperature thermopower provides with a very accurate measurement of the hole content in the $CuO_2$ planes. The value of S[290 K] was used to determine p following the study by Obertelli et al. [34]. As a double check, the parabolic $T_c(p)$ relation suggested by Presland et al. [35] was also used to estimate p. Both methods yield almost identical values of p agreeing within the limit ± 0.005. The SC transition temperatures of all the samples were determined from the $\rho(T)$ data (not shown). $T_c$ was located at the maximum of the $d\rho(T)/dT$ data. $T_c$ was also determined from the low-field ($H_{rms}$ = 1 Oe; f = 333.3 Hz) ACS (ac susceptibility) measurements for Y123 (not shown). A straight line was drawn on the steepest part of the diamagnetic ACS curve and another one was drawn as the T-independent base line representing the normal state ACS signal. The intercept between the two lines gave $T_c$. $T_c$ values obtained by these two different methods agree within ± 1.0 K for both the Y123 compounds considered here. We have taken the average value of these two $T_c$ values as the SC critical temperature for the analysis of fluctuation diamagnetism in this study. The SC transition temperatures are 92.1 K and 88.2 K for the p = 0.161 (optimum doping) and p = 0.143 (underdoped) compounds, respectively. It is interesting to note that these $T_c$ values are higher by ~ 1.5 K compared to the temperatures at which the maximum in the $d\chi(T)/dT$ occurs at a magnetic field of 5 Tesla.

*2.2. Measurement of magnetic susceptibility and results*

A *Quantum Design MPMS2* SQUID magnetometer was used for the measurement of $\chi(T)$ reported here, in the temperature range from 5 K to 400 K. A dc magnetic field of 5 Tesla was applied along the longest dimension of the sintered sample, with field-linearity checks within 1 Tesla – 5 Tesla (with a field interval of 1 Tesla) at 300 K and 100 K. Perfectly linear field dependence was found. The magnetic moment coming from the sample holder was measured and subtracted from the raw magnetization data. Data were collected following a predefined sequence. A scan length of 6 cm was used. The magnetic susceptibility was calculated from the mass of the samples and the background corrected magnetic moment. The magnetic susceptibilities of Y123 compounds are presented in Fig. 2.

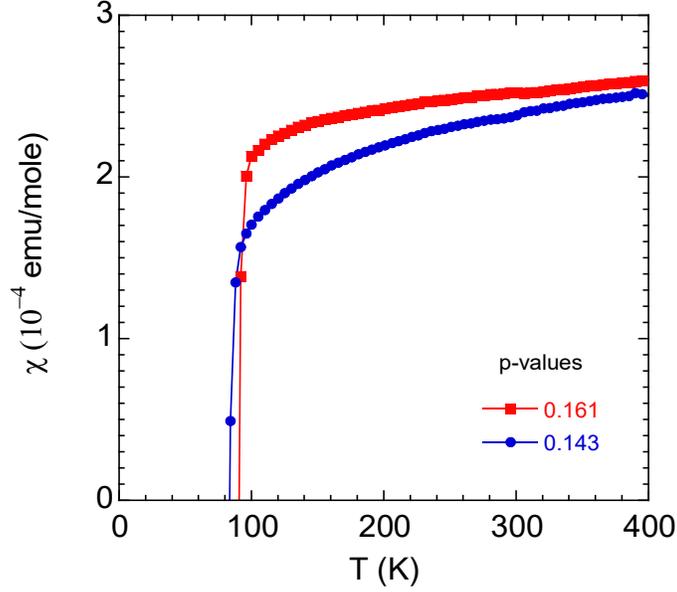

Figure 2: Temperature dependent magnetic susceptibility of Y123 compounds with different hole concentrations (p). For clarity some of the data points are not shown in the plot.

*2.3. Analysis of the fluctuation diamagnetism*

For reliable estimate of the contribution due to superconducting fluctuation to the magnetic susceptibility, an accurate fit to the normal state background susceptibility, $\chi_{bg}(T)$, is required. It is important to ensure that the fit to the susceptibility data in the normal state does not include any contribution originating from the fluctuating Cooper pairs. This essentially is a problem of setting the appropriate lower limit of temperature above which the $\chi(T)$ data would be used in the normal state to obtain $\chi_{bg}(T)$. The lowest temperature in the normal state at which the SC fluctuation is expected to vanish can be predicted from the consideration of the uncertainty principle and the minimum SC coherence length, $\xi(0)$ [29]. In fact Heisenberg uncertainty principle was applied to low-$T_c$ systems by Pippard to relate the size of the wave packet formed by the Cooper pairs, $\xi(0)$ to the $T_c$ [36, 37]. The suggestion made by Pippard states that the minimum size of a Cooper pair is of the order of $\xi(0)$; the SC coherence length at 0 K. If the finite temperature value of the coherence length is $\xi(T)$ - the characteristic length scale over which the density of Cooper pairs may vary, it may be concluded that even above $T_c$, where the phase incoherent Cooper pairs are created by thermal fluctuations, $\xi(T)$ must verify the condition, $\xi(T) \geq \xi(0)$. This condition, in terms of the reduced temperature [$\varepsilon = \ln(T/T_c)$], leads to $\varepsilon^c \sim 0.55$ as the thermal boundary above which SC fluctuations cease to exist. A value of $\varepsilon^c \sim 0.55$ implies that SC fluctuations are not expected to contribute significantly for temperatures above $1.7T_c$ [29]. We therefore, have used $\chi(T)$ data from $1.7T_c$ to 300 K to determine the background magnetic susceptibility of the Y123 compounds. The experimental $\chi(T)$ data has been fitted to a second order polynomial to obtain the background magnetic susceptibility. There is no strict and universally accepted procedure for fitting the normal state $\chi(T)$ devoid to SC contributions. For the compositions of Y123 under consideration, second order polynomial function fits the normal state $\chi(T)$ data in the temperature range $1.7T_c$ to 300 K very well. We have shown the results of these fits in Fig. 3. It is worth noticing that we have excluded high-T $\chi(T)$ data above 300 K for background fits. This is because the data at such high temperatures have little bearing on the possible diamagnetic SC fluctuations which are the focus of this work.

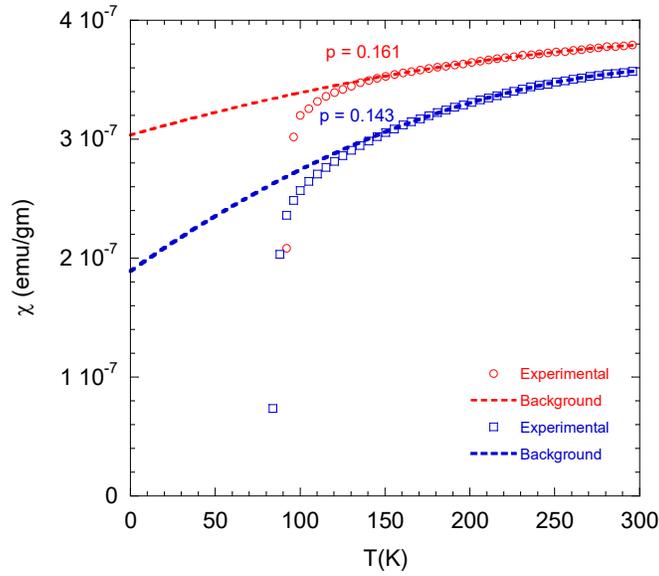

Figure 3: The experimental magnetic susceptibilities and respective background fits for Y123 compounds. For clarity some experimental data points are not shown.

It should be noted that, in Fig. 3, the molar magnetic susceptibility has been converted to mass magnetic susceptibility. Subsequent analysis of the fluctuation diamagnetism will be done keeping *cgs* units in mind. The fluctuation conductivity to the diamagnetic susceptibility, $\Delta\chi(T)$ is extracted from the expression $\Delta\chi(T) = \chi(T) - \chi(T)_{bg}$. Here, $\chi(T)$ is the experimental magnetic susceptibility and $\chi(T)_{bg}$ is the fit of the experimental magnetic susceptibility in the temperature range from 1.7$T_c$ to 300 K. To disclose the temperature dependence of the $\Delta\chi$ for the superconductors under consideration, we have shown $\Delta\chi(\varepsilon)$ for the compounds with p = 0.161 and p = 0.143 as a function of reduced temperature in Fig. 4. It is interesting to note that significant difference in $\Delta\chi(\varepsilon)$ for the two compositions sets in at quite high temperature ($\varepsilon \sim$ 0.50) even though the superconducting $T_c$ of the two compounds are quite close to each other.

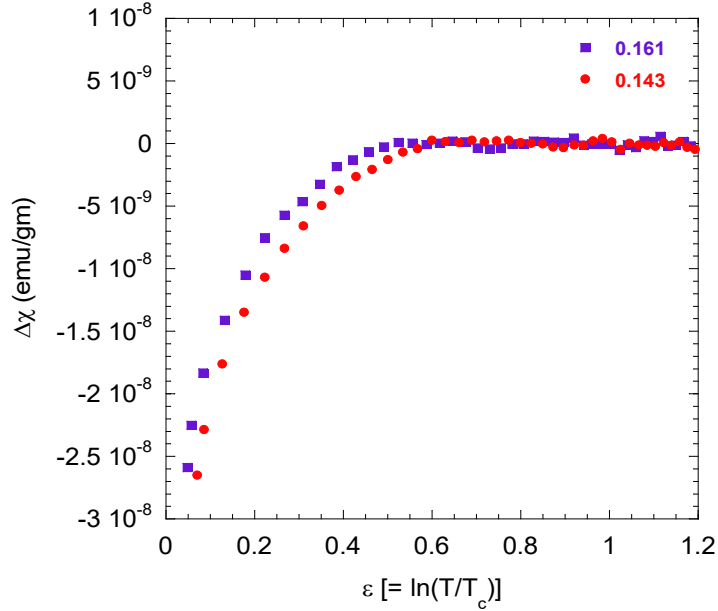

Figure 4: The fluctuation diamagnetic susceptibility versus reduced temperature of the two Y123 compounds. The hole contents are given in the plot.

By using the Ginzburg-Landau free energy functional, one can calculate the fluctuation diamagnetism of high-$T_c$ superconductors as a function of reduced temperature. In this work, we have adopted the methodology developed and employed by Rey et al. [30] to analyze the $\Delta\chi(T)$ data. The phenomenological mean field Gaussian Ginzburg-Landau formalism was used with appropriate energy cut-off parameter to suppress unrealistic high reduced temperature fluctuating modes. Within this approach, the fluctuation diamagnetic susceptibility is expressed as,

$$\frac{\Delta\chi(\varepsilon)}{T} = -\frac{\pi k_B \mu_0 \xi_{ab}^2(0)}{3\phi_0^2 s} \left[\frac{1}{\varepsilon}\left(1 + \frac{B_{LD}}{\varepsilon}\right)^{-1/2} - \frac{1}{\varepsilon^c}\left(1 + \frac{B_{LD}}{\varepsilon^c}\right)^{-1/2}\right] \quad (1)$$

where, $\mu_0$ is the vacuum permeability, $k_B$ is the Boltzmann constant, $\phi_0$ is the flux quantum, $\xi_{ab}(0)$ is the zero-T SC coherence length in the $CuO_2$ plane, s is the effective separation between the $CuO_2$ planes where fluctuating modes primarily reside, $B_{LD}$ is the Lawrence-Doniach parameter [4], and $\varepsilon^c$ is the dimensionless energy cut-off parameter. The Lawrence-Doniach parameter is given by,

$$B_{LD} = \left[\frac{2\xi_c(0)}{s}\right]^2$$

with $\xi_c(0)$ symbolizing the SC coherence length in the c-direction. At this point, it is worth noticing that Eqn. (1) is suitable for a situation where the magnetic field is applied parallel to the crystallographic c-direction (i.e., perpendicular to the $CuO_2$ planes). For sintered samples composed of polycrystalline aggregates, the average coherence length, $\xi(0) = \sqrt[3]{\xi_{ab}^2(0)\xi_c(0)}$, seems to be a more appropriate choice instead of $\xi_{ab}(0)$. We will address this issue further in Section 3. Selection of the value of s is a contentious issue. Two different assumptions are generally used [27, 30, 31, 38]. The periodicity in the $CuO_2$ planes has been taken either as 11.7 Å or 5.85 Å depending on the presumed level of coupling between the $CuO_2$ planes in the unit cell. In our analysis, both these values of s have been tried to check their effect on the quality of the fits to the fluctuation diamagnetic susceptibility data and also to explore their effect on the extracted values of $\xi_{ab}(0)$ and $\xi_c(0)$. We have found that use of s = 11.7 Å results in marginally better fits to the $\Delta\chi(T)/T$ data with reasonable values of the energy cut-off parameter. Moreover fits with s = 11.7 Å, yield values of $\xi_{ab}(0)$ and $\xi_c(0)$ close to those found in previous studies [31, 39 - 42]. Therefore, we display the fits for s = 11.7 Å only and confine our discussions with respect to these fits. The best fits to the $\Delta\chi(T)/T$ data are shown in Figs. 5. The second term in the square bracket of Eqn. (1) is absent when the effect of cut-off energy is not taken into account. We tried to fit the extracted $\Delta\chi(T)/T$ data without energy cut-off using the well-known Aslamazov-Larkin (AL) equation within the Lawrence-Doniach (LD) scenario. Such a scheme cannot fit the $\Delta\chi(T)/T$ data over the entire temperature range considered here with reasonable values of s, $\xi_{ab}(0)$ and $\xi_c(0)$.

Inspection of Figs. 5 reveals that fits to $\Delta\chi/T$ data for the optimally doped Y123 is quite good throughout the reduced temperature range considered, while for the underdoped Y123 there is significant deviation for $\varepsilon < 0.40$. The experimentally determined $\Delta\chi/T$ is higher in magnitude compared to the theoretically predicted fitted values for ε < 0.40 for the underdoped compound. This is the central result of this study.

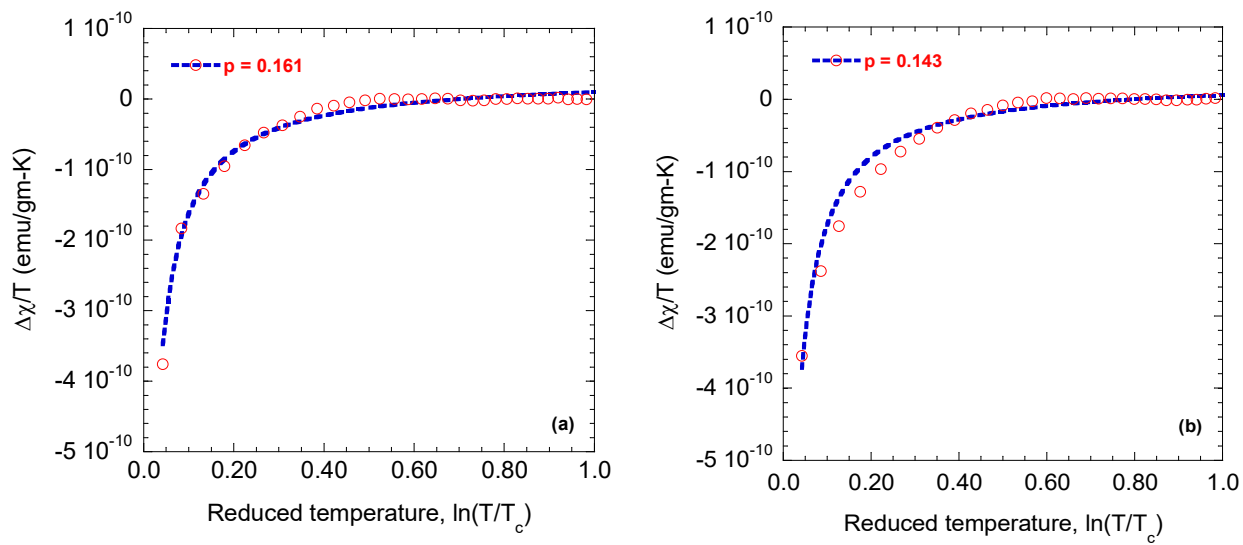

Figure 5: Fits to the Δχ/T data for the Y123 compounds. (a) p = 0.161 and (b) p = 0.143. The dashed lines are the fits to Eqn. (1). The extracted values of Δχ/T from the experimental magnetic susceptibilities are presented by the circles.

## 3. Discussion and conclusions

From the analysis of the fluctuation contribution to the diamagnetic susceptibility data, it is evident that Δχ/T for the underdoped compound cannot be modelled over the entire temperature range within the mean field Gaussian Ginzburg-Landau (MFGGL) scheme irrespective of theoretical details (with or without the energy cut-off condition). Before discussing the possible implication of these findings, clarification on a few points are in order.

Application of magnetic field is detrimental to superconductivity and can suppress SC fluctuations. The extent of the effect of the applied magnetic depends on the ratio (h) between the magnitudes of the applied field and the upper critical field $H_{c2}$; h ≡ $H/H_{c2}$. The magnetic susceptibility data used for the analysis of fluctuation diamagnetism was obtained with an applied magnetic field of 5 Tesla. The analysis of Δχ(ε)/T [Eqn. (1)] was performed in the zero magnetic field limit (the Schmidt limit) [30]. A question naturally arises, whether this assumption is valid or not? The Schmidt limit works when h/ε < 1. The values of the upper critical fields for Y123 vary over a wide range within 100 Tesla to 400 Tesla in the literature [41 – 45]. For both the compositions of Y123 considered in this study, superconducting transition temperatures are high, close to that at the optimum doping. The upper critical fields are also high [41 – 45]. If we take $H_{c2}$ ~ 200 Tesla, h = 0.025. The lower limit of the reduced temperature has been set around ~ 0.05 for the fit of fluctuation contribution to the diamagnetic susceptibility. This yields h/ε ~ 0.5 and implies that the analysis within the Schmidt limit is largely valid for most of the temperature region used in this study. There are two main considerations for selecting the lower limit of the reduced temperature ~0.05. One is the validity of the method of analysis and the other is related to the critical fluctuations. For SC cuprates, the coherence lengths are very small compared to the conventional low-$T_c$ systems. Therefore, for cuprates the temperature width of critical fluctuations above the superconducting transition temperatures are greatly enhanced [46]. At temperatures too close to the mean-field $T_c$, critical fluctuations dominate and the MFGGL treatment breaks down. The extracted values of SC coherence lengths from the best fits to the Δχ/T data are as follows: ξ(0) = 25 Å and $ξ_c(0)$ = 1.3 Å for p = 0.161 and ξ(0) = 26 Å and $ξ_c(0)$ 1.5 Å for the p = 0.143 compound. The extracted values of $ξ_c(0)$ are in very good agreement with those reported in earlier studies [30, 31, 39 – 42]. The values of ξ(0) (Section 2) are close to the reported values of $ξ_{ab}(0)$ for Y123 [30, 31, 39 – 42]. It is interesting to note that, magnetization measurements were done on randomly oriented superconducting grains in the sintered polycrystalline compounds. The closeness between ξ(0) and $ξ_{ab}(0)$ indicates that the fluctuating screening supercurrent, giving rise to fluctuation diamagnetism, flows mainly within the $CuO_2$ planes. The values of cut-off parameter ($ε^c$) were 0.73 and 0.81 for p = 0.161 and p = 0.143 compounds; in perfect agreement with earlier investigations [29].

Careful inspection of Fig. 5(b) reveals that experimental Δχ/T falls below the fitted line in the reduced temperature range from ~0.40 to ~0.20 where the PG induced downturn in χ(T) is expected to be significant in the p = 0.143 compound [12]. The agreement between the theoretical prediction and experimental Δχ/T is much better for the optimally doped compound with p = 0.161 [Fig. 5(a)], where the effect of the PG on χ(T) is largely reduced. This notable difference between the two fits in describing the Δχ/T data implies that PG induced reduction in the magnetic susceptibility is not related directly to the superconducting fluctuations and cannot be described within the MFGGL formalism. These findings suggest that PG and SC pairing correlations probably have different physical origins. Similar conclusions were drawn from the analysis of paraconductivity data of Y123 with different hole contents [31].

**Acknowledgements**

R.S.I. and S.H.N. acknowledge the research grant (1151/5/52/RU/Science-07/19-20) from the Faculty of Science, University of Rajshahi, Bangladesh, which partly supported this work.

**Data availability**

The data sets generated and/or analyzed in this study are available from the corresponding author on reasonable request.

**Author Contributions**

**Additional Information**

**Competing Interests**

The authors declare that they have no known competing financial interests or personal relationships that could have appeared to influence the work reported in this paper.